# Solitons in the nonlinear Schrödinger equation with the pseudo-Raman term, anti-diffusion gain and nonlinear loss

**E. M. Gromov**[1, *] **and B. A. Malomed**[2,3]

[1]National Research University Higher School of Economics, Nizhny Novgorod 603155, Russia

[2]Department of Physical Electronics, School of Electrical and Computer Engineering, Faculty of Engineering, and Center for Light-Matter Interaction, Tel Aviv University, Tel Aviv 69978, Israel

[3]Instituto de Alta Investigación, Universidad de Tarapacá, Casilla 7D, Arica, Chile

## Abstract

We introduce a nonconservative nonlinear Schrödinger equation (NLSE) which includes additional nonlinear terms accounting for the pseudo-stimulated-Raman-scattering (pseudo-SRS) effect and cubic loss. An essentially novel feature of the model is the linear anti-diffusion gain, which leads to the obvious blow-up in the infinite or spatially periodic system. We consider the absorptive boundary conditions, which suppress the blow-up driven by effective convective instability in the setup. The NSLE model is derived, in the extended adiabatic approximation, from the underlying Zakharov's system, which includes the viscosity term acting on its low-frequency component and the anti-diffusion gain in the high-frequency (HF) one. The latter term appears if the expansion, with respect to the deviation of the wavenumber, is performed not around the minimum-loss point (as in the usual situation), but around a maximum-losspoint. Dynamics of moving solitons in the model is addressed by means of analytical and numerical methods. The sech approximation for the soliton predicts it as a stable fixed point (FP) of the system of balance equations for the HF norm and wave momentum, derived in the moving reference frame. The FP's attraction basin is identified too. The predictions are corroborated by direct simulations of the full NSLE with the absorptive boundary conditions.

## Highlights

>Solitons in the nonconservative nonlinear Schrodinger equation (NSLE) are considered. > The NSLE includes the pseudo-Raman effect, nonlinear loss, and anti-diffusion gain. >The NSLE is derived from the Zakharov's system for coupled high- and low-frequency components. >The soliton`s dynamics is investigated by means of analytical and numerical methods.

* Corresponding author. *E-mail address:* egromov@hse.ru (E.M. Gromov).

**Keywords**: Nonconservative Nonlinear Schrodinger Equation; Stimulated Raman Scattering; Nonlinear Loss, Anti-diffusion Gain; Convective Instability; Fixed Point; Solitons

## 1. Introduction

Solitons are robust localized modes which are maintained in a great variety of physical realizations by stable balance of nonlinearity and linear dispersion or diffraction [1-4]. These realizations may be provided by settings demonstrating the propagation of high-frequency (HF) and/or low-frequency (LF) waves. In the former case, solitons appear as broad self-trapped envelopes of HF oscillations, with well-known examples provided by solitons in optical media with the Kerr nonlinearity [1-3] (which may include left-handed metamaterials [5-9]), Langmuir solitons in plasmas [10-12], matter-wave solitons in ultracold bosonic gases [13-16], nonlinear surface-wave excitations on deep water [17-19], and Davydov solitons in long molecules [20,21]. The nonlinear Schrödinger equation (NLSE) provides a universal model for envelope solitons [22,23]. On the other hand, nonlinear LF fields may directly self-trap into solitons, such as those built by ion-acoustic [24] and magneto-acoustic [25] waves in plasmas, and surface waves in shallow water and long internal waves in stratified fluids [26,27]. The fundamental model for nonlinear LF media is provided by the Korteweg – de Vries equation (KdVE) and its generalizations.

More complex systems often include coupling between co-propagating HF and LF waves, whose frequencies differ by more than an order of magnitude, while group velocities are equal or close. In plasmas, the ponderomotive force gives rise to the nonlinear coupling between the HF Langmuir and LF ion-acoustic waves [10], which is modeled by the celebrated Zakharov's system (ZS) [10]. In the ocean, HF waves are surface modes, hydrodynamically coupled to LF internal waves propagating in deep stratified water [17-19]. In long molecules, the HF and LF excitations are represented by the intrinsic vibrations and acoustic waves, respectively [20,21].

The ZS is a universal system modeling the HF-LF interactions in diverse settings, including the above-mentioned ones. Unlike the NLSE and KdVE in their basic forms, the ZS is not an integrable model [28]. Nevertheless, it gives rise to stable single-soliton solutions. Other well-elaborated implementations of the ZS represent the interplay of HF oscillations of atomic positions and acoustic waves in long molecular chains, which may create Davydov solitons [29], and the coupling of HF surface and LF internal waves in stratified fluids [30].

The ZS may be simplified, replacing the second-order bidirectional LF equation by one reduced for the unidirectional propagation. In particular, KdVE may naturally play the role of the LF equation, being coupled to the NLSE for the HF waves, see Eqs. (1) and (2) below. The LF component may be eliminated in the adiabatic approximation, in which the slowly evolving LF field is replaced by a term representing the local density of the HF component. In the adiabatic approximation, the ZS reduces to the NLSE.

In many cases a realistic model includes viscous damping of the LF wave. In the presence of the damping, the adiabatic approximation still makes it possible to replace the full ZS by the single-component NLSE, which includes a dissipative nonlinear term ~ $\mu$ in Eq. (3) (see below). The same term in the NLSE with time $t$ and coordinate $x$ replaced, respectively, by the propagation distance, $z$, and "retarded time", $\tau$, represents the effect of the stimulated Raman scattering (SRS) in fiber optics [31-33], i.e., splitting of the pump photon into a longer-wavelength secondary one (the *Stokes wave*) and an optical phonon (the *idler*, which represents an effective loss, in terms of optical fields). Phenomenologically, the SRS effect in fibers is a manifestation of a small delay (~ a few fs) of the nonlinear response of silica to the temporal variation of the optical field [1,3,32]. On the contrary to the time-domain SRS in fiber optics, the formally similar term in Eq. (3) emulates the SRS effect in the spatial domain. Therefore, it was called a *pseudo-SRS* effect in works [34-38], where it was derived, as said above, by means of the adiabatic approximation applied to the ZS, including viscosity acting in the LF component. The pseudo-SRS effect was studied in the context of the soliton dynamics [34-37], as well as in terms of the modulational instability of continuous waves [38]. A common feature of the temporal-domain SRS and spatial pseudo-SRS effects is their dissipative nature.

The dynamics of narrow temporal solitons under the action of the SRS effect in fiber optics was studied in detail experimentally [31] and theoretically [32,33]. In particular, it is well known that SRS shifts the soliton's spectrum towards longer wavelengths, which may be interpreted as an effective self-deceleration of the temporal solitons, eventually leading to their destruction. This trend may be compensated by the use of a frequency-sliding gain [39], which is applied to compensate losses in long-haul fiber-optic communication lines [40]. Other possibilities for the suppression of the SRS-induced drift of the spectrum are offered by the emission of small-amplitude waves from the temporal soliton [41], periodic modulation of the group-velocity dispersion [42,43], shift of the zero-dispersion point [44], and the use of dispersion-decreasing fibers [45]. In addition to inducing the self-deceleration of the solitons, SRS produces a similar effect on optical shock waves [46,47]

Possibilities for stabilization of spatial solitons against the action of the pseudo-SRS effect were analyzed too [34-37,48,49]. In particular, it was demonstrated that the stabilization may be provided by variable dispersion [34-37] or by an external potential, which may be naturally added to the effective NLSE in the contexts of the underlying physical realizations in the spatial domain [48]. A related model is based on an extended Lugiato-Lefever equation with the pseudo-SRS cubic term, linear damping, and spatially inhomogeneous pump [49].

Realistic models of fiber lasers and telecommunication fiber links include gain, loss, group-velocity dispersion, and nonlinear effects [50-54]. The wave propagation in the gain fiber is governed by the cubic complex Ginzburg-Landau equation with the gain term subject to saturation and limited bandwidth [55-60]. Further relevant results are reported in recent studies of the existence, bifurcations, and stability of nonlinear localized states [61,62].

Another relevant possibility, which was not addressed previously, and is considered in the pre present work, is that the physical settings modeled by the nonconservative NSLE (with the pseudo-SRS and cubic-loss terms) includes, on the contrary to the usual models, linear gain of the *anti-diffusion* type, represented by the second-derivative term with the sign opposite to that which represents diffusion/spectral-filtering losses in the usual settings. The source of the effective diffusion (viscosity) is well known: the effective NLSE is derived assuming that the underlying carrier wavenumber corresponds to the *minimum linear loss* in the system; then, the effective diffusion term represents the increase of the loss strength with deviation of the wavenumber from the minimum [1]. We consider the opposite situation, with the carrier wavenumber corresponding to the *maximum* of the linear loss (the maximum may be imposed, in particular, by resonant absorption). Then, the decrease of the loss strength corresponding to the deviation of the wavenumber from the maximum point is modeled by the anti-diffusion term, in the framework of the nonconservative NLSE. Obviously, such a gain term would lead to the blow-up in the infinite or spatially periodic system. However, we consider the setup of a finite size with absorptive boundary conditions at its edges. The combined action of the linear dispersion and SRS term makes growing perturbations, pumped by the anti-diffusion term, moving, i.e., the gain-driven instability is *convective*. As a result, growing perturbations eventually hit the system's edge, where they are suppressed by the boundary absorption. As a result, the model supports *stable solitons*, which we investigate by means of combined analytical and numerical methods.

The subsequent presentation is organized as follows. The model is formulated in Section 2, and approximate analytical results for solitons are reported in Section 3. Using evolution equations for the norm and wave momentum, applied to the sech ansatz, we predict a soliton solution as a fixed point (FP) of the respective dynamical system, along with its attraction basin. In Section 4, the predictions of the analytical approximation for the soliton are corroborated by direct simulations of the model equation with the absorptive boundary conditions (BCs). The paper is concluded by Section 5.

## 2. The model

We consider the evolution of a slowly varying envelope $\Psi(x,t)$ of the HF wave field (e.g., the local amplitude of Langmuir waves in a plasma) in the one-dimensional medium with the cubic self-spiraling nonlinearity, under the action of anti-diffusion gain with coefficient $p > 0$. The HF equation, which includes the cubic loss with coefficient, is nonlinearly coupled to the unidirectional equation for the LF field, $n(x,t)$ (e.g., local perturbation of the ion density of the plasma), including the usual viscosity/diffusion term with coefficient $\mu > 0$. As a result, we arrive at the following generalized ZS:

$$2i\frac{\partial \Psi}{\partial t} + \frac{\partial^2 \Psi}{\partial x^2} - 2n\Psi + ip\frac{\partial^2 \Psi}{\partial x^2} + ig\Psi|\Psi|^2 = 0\,, \qquad (1)$$

$$\frac{\partial n}{\partial t} + \frac{\partial n}{\partial x} - \frac{\mu}{2}\frac{\partial^2 n}{\partial x^2} = -\frac{\partial\left(|\Psi|^2\right)}{\partial x}\,, \qquad (2)$$

where the coefficients in front of the second and fourth terms in Eq. (2) are fixed to be 1 by means of scaling.

To understand the result of the action of the anti-diffusion gain in Eq. (1), we represent the solution the respective linearized equation (1) for $n = 0$ as $\Psi \sim \exp(ikx - i\omega t)$, with the obvious dispersion relation, $\omega = k^2/2 + ipk^2/2$. The real part of the frequency determines the group velocity of the HF field, $V_{\mathrm{gr}} = d(\mathrm{Re}\,\omega)/dk = k$. The growth the HF field (the instability) is determined by the imaginary part of the frequency, $\Psi \sim \exp(\mathrm{Im}\,\omega \cdot t) \equiv \exp\left(pk^2 t/2\right)$.

Solutions of Eq. (2) for small-amplitude LF waves take the form of $n \sim \exp(iKx - i\Omega t)$, with the respective dispersion relation $\Omega = K - i(\mu/2)K^2$. The real part of frequency $\Omega$ accounts for the propagation of the LF wave field at a speed of $V_{LF} = (\mathrm{Re}\,\Omega)/K \equiv 1$, while the imaginary part of $\Omega$, with $\mu > 0$ implies the attenuation or growth of the LF component, $n \sim \exp(\mathrm{Im}\,\Omega t) \equiv \exp\left(-\mu K^2 t/2\right)$.

In the lowest adiabatic approximation, which neglects the derivative terms $\partial n/\partial t$, and $(\mu/2)\partial^2 n/\partial x^2$ in Eq. (2), it is reduced to a local relation,

$$n = -|\Psi|^2\,. \qquad (3)$$

Accordingly, Eq. (1) is replaced by the nonconservative NSLE with the anti-diffusion gain:

$$2i\frac{\partial \Psi}{\partial t} + \frac{\partial^2 \Psi}{\partial x^2} + 2\Psi|\Psi|^2 + ip\frac{\partial^2 \Psi}{\partial x^2} = 0\,. \qquad (4)$$

In the next approximation, the viscosity term in Eq. (2), which was neglected at the previous stage, is taken into account by the substitution of relation (3): $(\mu/2)\partial^2 n/\partial x^2 \approx -(\mu/2)\partial^2\left(|\Psi|^2\right)/\partial x^2$, to produce an amended local relation, $n = -|\Psi|^2 - (\mu/2)\partial\left(|\Psi|^2\right)/\partial x$ and, consequently, the following nonconservative NSLE:

$$2i\frac{\partial\Psi}{\partial t} + \frac{\partial^2\Psi}{\partial x^2} - 2\Psi|\Psi|^2 + ip\frac{\partial^2\Psi}{\partial x^2} + ig\Psi|\Psi|^2 + \mu\Psi\frac{\partial\Psi\left(|\Psi|^2\right)}{\partial x} = 0. \quad (5)$$

Here the term $\sim\mu$ represents the above-mentioned pseudo-SRS effect in the spatial domain, and as said above, $g$ and $p$ are the parameters of the cubic loss anti-diffusion gain.

The linearized form of Eq. (5) is

$$2i\frac{\partial\Psi}{\partial t} + \frac{\partial^2\Psi}{\partial x^2} + ip\frac{\partial^2\Psi}{\partial x^2} = 0. \quad (6)$$

This equation has a solution $\Psi(x,t) = \Psi_0 \exp\left(ikx - i\omega t + pk^2 t/2\right)$ with $\omega = k^2/2$, which is exponentially unstable for $p > 0$. These are waves traveling with group velocity $V_{\mathrm{gr}} = d(\mathrm{Re}\,\omega)/dk = k$, and exponentially growing as $\exp\left(pk^2 t/2\right)$. In the infinite system or a system with periodic BCs, the exponential growth will lead to the blow-up. To prevent it, we use dissipative BCs in the form of a spatially inhomogeneous parameter $p(x)$, which changes from positive to negative values:

$$p \to p(x) = p_0 \tanh\left(L - |x|\right). \quad (7)$$

In this case, in the region of $|x| < L$, one has an unstable medium with the convective instability, while the lossy (viscous) medium absorbs the propagating disturbances at $|x| > L$, which eventually leads to the suppression of the blow-up.

As a characteristic example, we consider the evolution of the initial wave packet

$$\Psi(x,0) = \mathrm{sech}(x/5)\exp(i5x) \quad (8)$$

in the framework of the linearized equation (5) with the absorbing boundary regions introduced as per Eq. (7):

$$2i\frac{\partial\Psi}{\partial t} + \frac{\partial^2\Psi}{\partial x^2} + ip_0 \tanh\left(L - |x|\right)\frac{\partial^2\Psi}{\partial x^2} = 0, \quad (9)$$

for $L = 20$ and different values $p_0$. Figures 1(a) and 1(b) show the spatiotemporal distribution of $|\Psi(x,t)|$ produced by direct numerical simulations of Eq. (9) with $p_0 = 1/40$ and $p_0 = 1/20$, respectively It is clearly seen that no waves are reflected from the boundary, and $|\Psi(x,t)|$ attains a maximum in the region where the parameter $p(x)$ crosses zero as per Eq. (7).

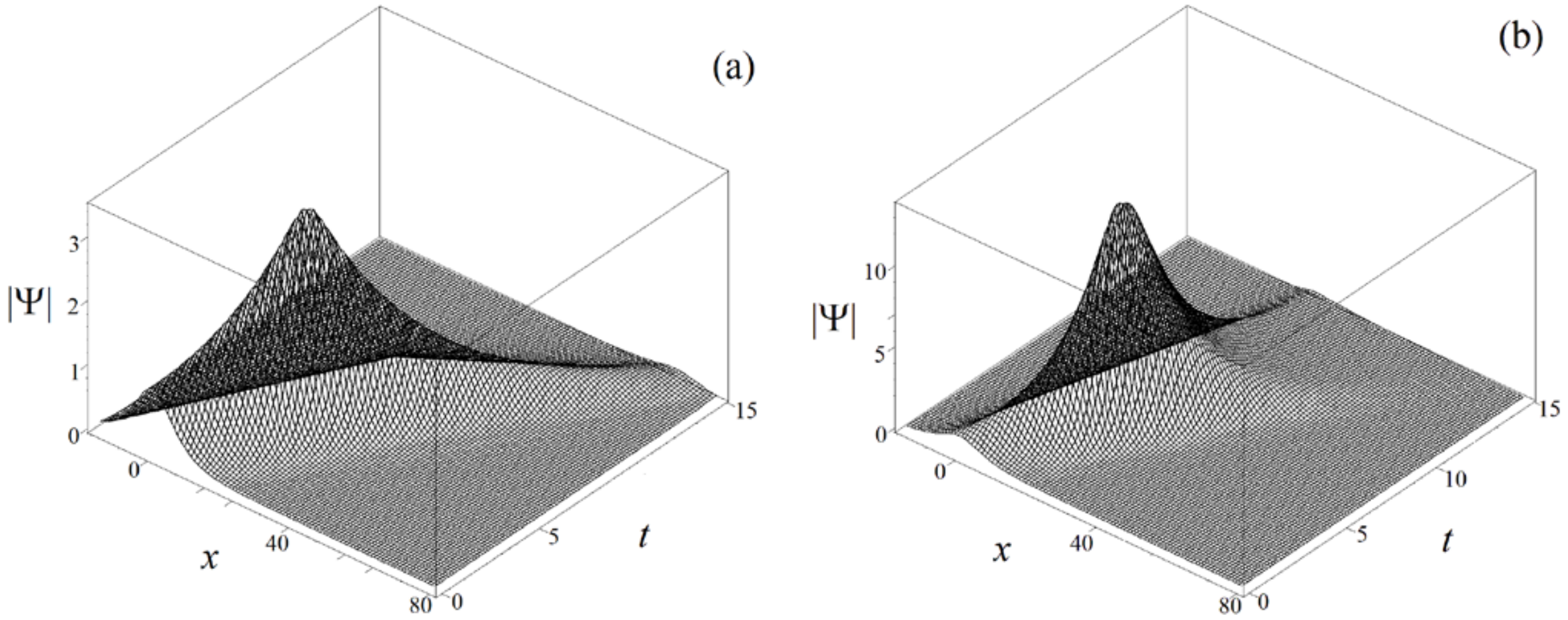

**Fig. 1.** The spatiotemporal distribution $|\Psi(x,t)|$ produced by the simulations of the linearized equation (9) with the initial distribution (8) and $p_0 = 1/40$ (a) or $p_0 = 1/20$ (b).

Figure 2 additionally illustrates the suppression of the blow=up by displaying the maximum values of $|\Psi(x,t)|$ vs. time for the same cases $p_0$.

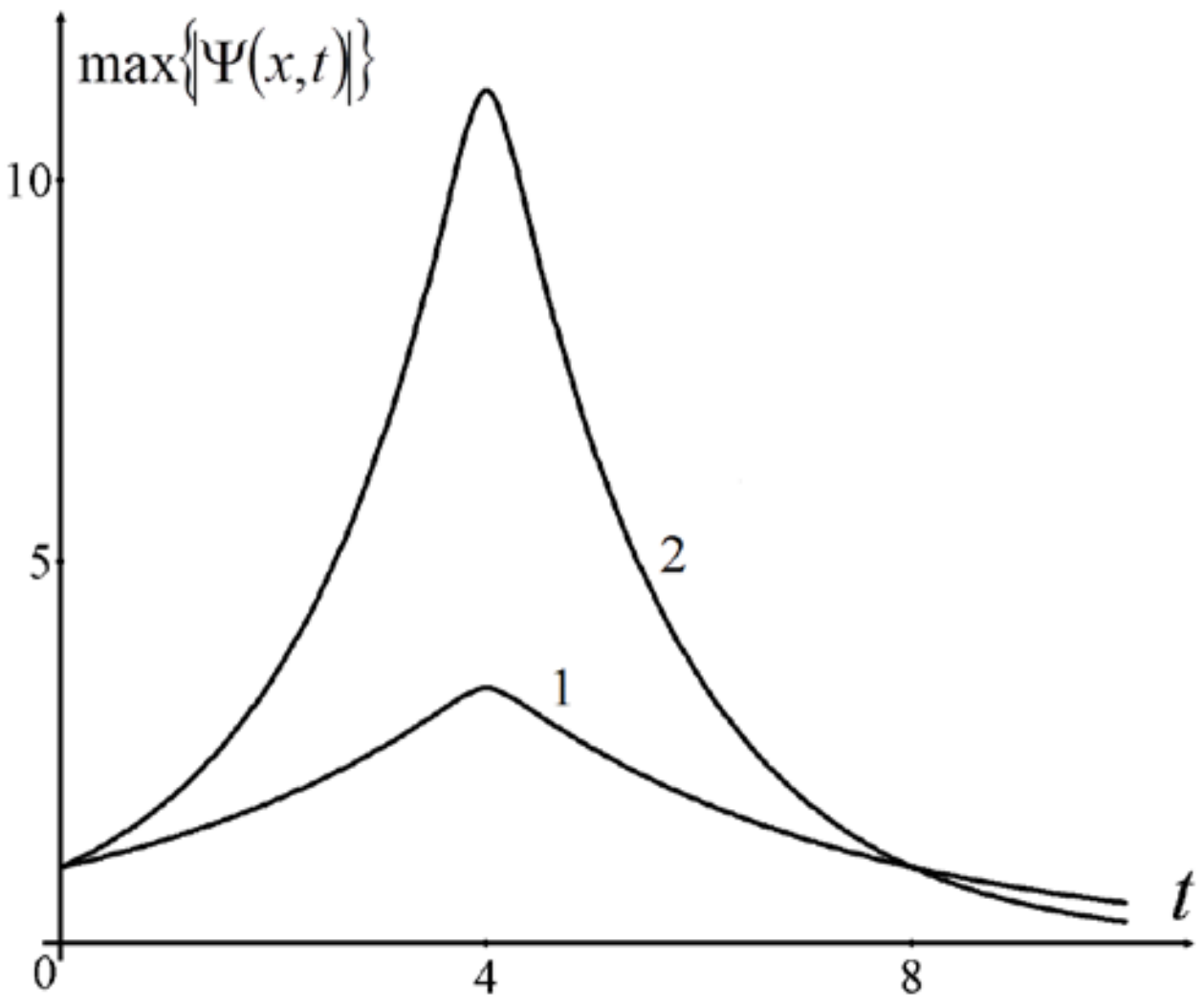

**Fig. 2.** The maximum value of $|\Psi(x,t)|$ vs. time, as produced by the same numerical simulations which are presented in Fig. 1.

For analytical assessment of the dynamics of wave packets in the framework of Eq. (9), we approximate them by the ansatz with amplitude $A$ and spatial width $\Delta$,

$$\Psi(x,t) = A(t)\operatorname{sech}[(x - Vt)/\Delta]\exp(ikx - i\omega t). \tag{10}$$

In the reference system moving with speed of the wave packet, $\xi = x - Vt$, setting $|x| = Vt$, for $k\Delta >> 1$ we obtain an approximate equation for the evolution of the amplitude:

$$2\frac{dA}{dt} + Ap_0 \tanh(L - Vt) = 0 \,. \tag{11}$$

The solution of Eq. (11) is

$$A(t) = A(0)\left[\frac{\cosh(L - Vt)}{\cosh L}\right]^{-p_0 k/2} . \tag{12}$$

According to this approximation, the packet`s amplitude attains its maximum, $A(t) = A(0)(\cosh(L - Vt))^{p_0 k/2}$, at the moment at time $t_* = L/V$, when $dA(t_*)/dt = 0$. Thus, for $p_0 = 1/40$, $k = 5$ и $L = 20$, we obtain the maximum value $A \approx 3.5$, and $A \approx 15$ for $p_0 = 1/20$, in approximate agreement with the numerical results presented in Fig. 2. Overall, solution (12) matches the results of the numerical simulations quite well.

In the analytical consideration of the full nonlinear model, which is presented below in Section 3, we address Eq. (5) with a constant value of $p$. This assumption is relevant for the inner part of the system, where $p$ remains approximately constant. Numerical simulations of the nonlinear model are reported in Section 4 in the framework of Eq. (27), with the spatially inhomogeneous parameter $p(x)$ defined as per Eq. (8).

## 3. Analytical approximations

## 3.1. Integral relations

The evolution equations for naturally defined integral moments of the HF field are easily derived from Eq. (5) for spatially localized wave packets, so that $\Psi|_{x\to\pm\infty} \to 0$:

$$\frac{dP}{dt} \equiv \frac{d}{dt}\int_{-\infty}^{\infty} k|\Psi|^2 dx = -\frac{\mu}{2}\int_{-\infty}^{+\infty}\left[\frac{\partial\left(|\Psi|^2\right)}{\partial x}\right]^2 dx - ip\int_{-\infty}^{+\infty}\frac{\partial^2\Psi}{\partial x^2}\frac{\partial\left(\Psi^*\right)}{\partial x}dx - gM \,, \tag{13}$$

$$\frac{dN}{dt} \equiv \frac{d}{dt}\int_{-\infty}^{\infty}|\Psi|^2 dx = p\int_{-\infty}^{+\infty}\left|\frac{\partial\Psi}{\partial x}\right|^2 dx - gM \,, \tag{14}$$

where $M = \int_{-\infty}^{\infty}|\Psi|^4 dx$. Here, we use the Madelung representation $\Psi = |\Psi|\exp(i\vartheta)$, with $k = \partial\vartheta/\partial x$ being the local wavenumber. Obviously, $N$ and $P$ define the total norm and wave momentum of the wave packet.

### 3.2. Effective evolution equations

To analyze the wave-packet dynamics in the framework of Eqs. (13)-(14), we adopt the usual sech ansatz for solutions, with variable amplitude $A(t)$, and wavenumber $k(t)$ (cf. Eq. (10)),

$$\Psi(x,t) = A(t)\operatorname{sech}\left[A(t)\left(x - \int k(t)dt\right)\right]\exp\left[ik(t)x + i(1/2)\int A^2(t)dt\right]. \tag{15}$$

Substituting the ansatz (15) in Eqs. (13)-(14), we derive the following system of evolution equations for the time-dependent wavenumber and amplitude:

$$\frac{dk}{dt} = -\frac{4}{15}\mu A^4 + \frac{2}{3}pkA^2, \tag{16}$$

$$\frac{dA}{dt} = \frac{(p-2g)}{3}A^3 + pk^2 A. \tag{17}$$

**3.3. Equilibrium states (fixed points) of the dynamical system (16)-(17).**

To analyze the solutions, we use substitutions

$$\chi = k\mu / p,\ a = (A\mu / p)^2 \text{ and } \tau = tp^3 / \mu^2, \tag{18}$$

which reduces the system (16)-(17) to the scaled form with the single free parameter, $\sigma \equiv 2(g/p) - 1$:

$$\frac{d\chi}{d\tau} = -\frac{4}{15}a^2 + \frac{2}{3}\chi a, \tag{19}$$

$$\frac{da}{d\tau} = -\frac{2}{3}\sigma a^2 + 2\chi^2 a. \tag{20}$$

Straightforward analysis of Eqs. (19)-(20) demonstrates that they give rise to the following stationary (equilibrium) state, alias a fixed point (FP):

$$a_* = 25\sigma/12 > 0,\quad \chi_* = 5\sigma/6 \ . \tag{21}$$

Straightforward analysis of small perturbations around the FP, governed by Eqs. (20) and (21), demonstrates that it is stable at $\sigma \geq 1$. In terms of the original notation, the FP coordinates, i.e., equilibrium values of the soliton's amplitude and wavenumber, are

$$A_*^2 = \frac{25}{12}\frac{p^2}{\mu^2}\left(2\frac{g}{p} - 1\right) > 0,\quad k_* = \frac{2}{5}\frac{\mu}{p}A_*^2 \equiv \frac{5}{6}\frac{p}{\mu}\left(2\frac{g}{p} - 1\right). \tag{22}$$

Note that the FP represents a state steadily moving with speed $k_*$.

**3.4. Time-dependent solutions of the dynamical system (19)-(20).**

For $\sigma = 1$, which corresponds to the equality of the nonlinear-loss and anti-diffusion-gain coefficients ($g = p$, in the original notation), the FP (22) amounts to

$$\chi_* = 5/6 \approx 0.83, \; a_* = 25/12 \approx 2.08. \tag{23}$$

It is an FP of the center type, with the eigenfrequency of small oscillatory perturbations $\omega = 25/18$. In this case, the system of Eqs. (19)-(20) admits an integral of motion,

$$a^2 - 5a\chi + 5\chi^3 = C \equiv \text{const}. \tag{24}$$

Figure 3 shows phase trajectories (24) for different values of $C$. The FP (23) corresponds to $C = C_* \equiv -(25/12)^2/3 \approx -1.45$, while values $C_* < C < 0$ and $C > 0$ correspond to closed and open trajectories, respectively, with the boundary between them represented by the separatrix corresponding to $C$ = 0 (curve 2 in Fig. 3). In particular, curves 1 and 3 in Fig. 3 are plotted for $C = -1$ and $C = +2.5$, respectively.

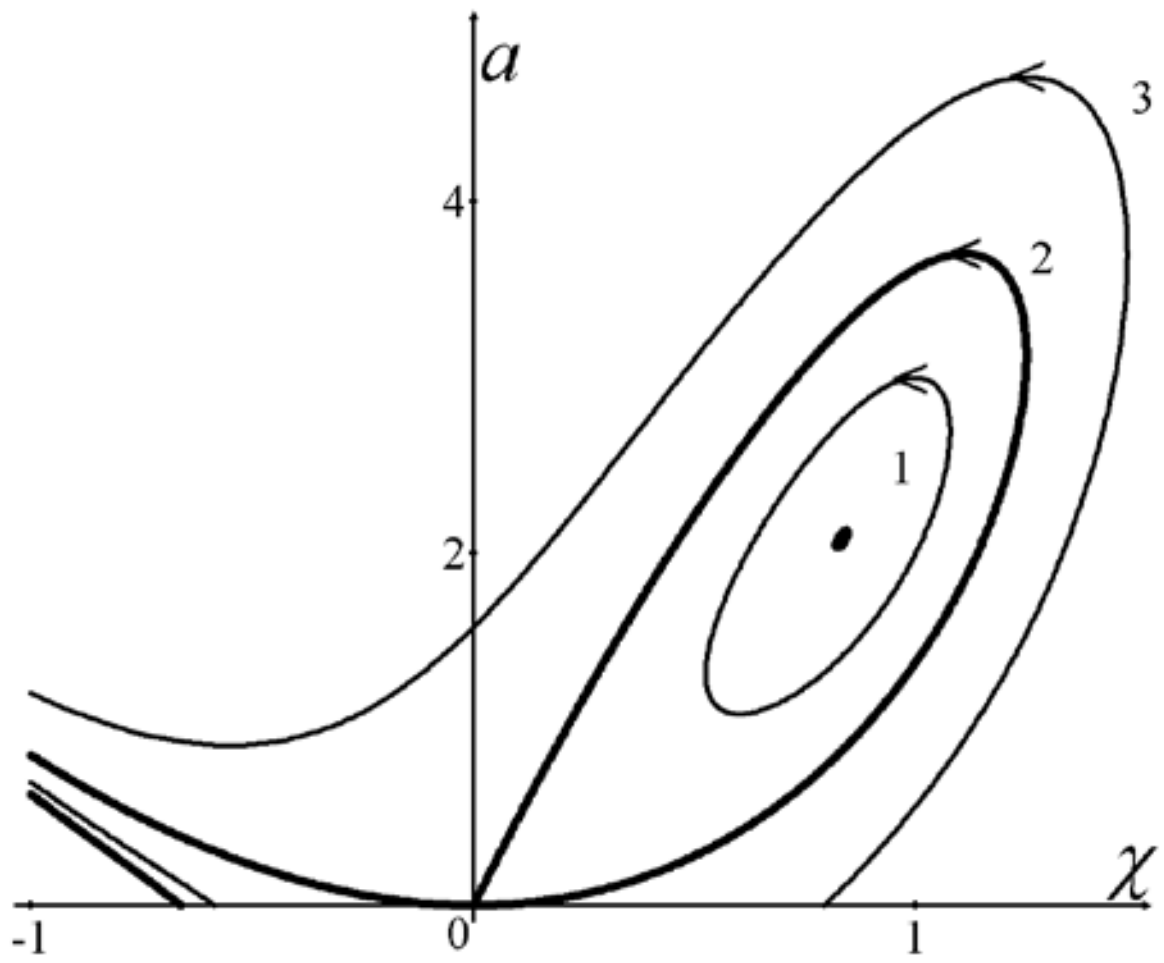

**Fig. 3.** Phase trajectories (24) for different values of $C$. Here and in Fig. 4, arrows indicate the direction of motion along the trajectories. The FP $\chi_* = 5/6 \approx 0.83$, $a_* = 25/12 \approx 2.08$ corresponds to $C = C_* \equiv -(25/12)^2/3 \approx -1.45$, curves 1 and 3 correspond to $C = -1$ and $C = +2.5$, respectively, while the separatrix 2 pertains to $C = C_{\text{cr}} = 0$.

In particular, solutions of the dynamical system (19), (20) with initial conditions $(\chi(0) = \chi_* = 5/6, a(0))$ correspond to $C_* < C < 0$ and thus produce closed trajectories in the interval of initial intensities

$$0.42 \approx 1 - 1/\sqrt{3} < a(0)/a_* < 1 + 1/\sqrt{3} \approx 1.58. \tag{25}$$

For $\sigma > 1$ the FP (21) of the system (19)-(20) is a stable spiral. For instance, consider the system with $\sigma = 3/2$. In this case, the FP (22) has coordinates $a_* = 25/8$, $\chi_* = 5/4$. Figure 4 shows phase trajectories as a result of numerical simulations of the system for this case, with the initial wavenumber corresponding to the equilibrium value $\chi(0) = \chi_* \equiv 5/4$, and various initial values of intensity $a(0)$. Curves 1, 2, 3, and 4 correspond, respectively, to $a(0) = 0.01a_*$, $a(0) = 1.9a_*$, $a(0) = 2.05a_*$, and $a(0) = 2.2a_*$. Curves 1 and 2 represent the oscillatory motion attracted to the FP, while curve 3 is pulled towards zero. The interval of initial intensity values $a(0)$ in which the attraction to the FP occurs is

$$0 < a(0)/a_* \le 2. \tag{26}$$

For $a(0)/a_* > 2$, solutions do not belong to the attraction basin of the FP and evolve towards $a = 0$ (e.g., curves 3 and 4 in Fig. 4), which implies the gradual decay of the input pulse (15).

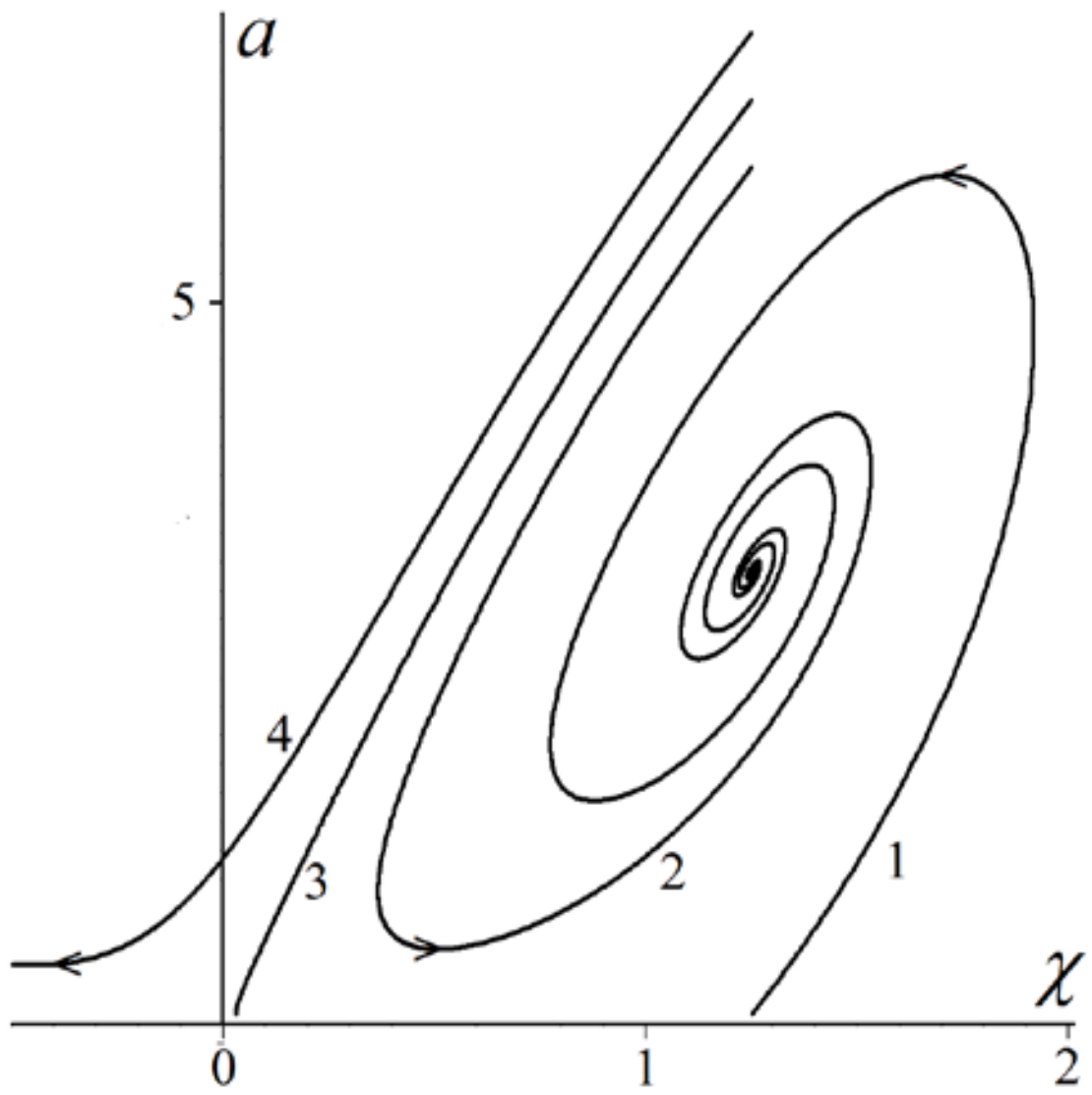

**Fig. 4.** Phase-plane trajectories of numerical solutions of Eqs. (19)-(20) with $\sigma = 3/2$, the initial value $\chi(0) = \chi_* \equiv 5/4$ and different initial values of $a(0)$: curve 1 for $a(0) = 0.01a_*$, curve 2 for $a(0) = 1.9a_*$, curve 3 for $a(0) = 2.05a_*$, and curve 4 for $a(0) = 2.2a_*$.

## 4. Results of full numerical simulations

As said above, full numerical simulations were carried out for the nonconservative NLSE (4) with the constant anti-diffusion gain coefficient *p* replaced by the spatially modulated one (7), which accounts for the absorption at the boundaries. Thus, the model equation for numerical modeling was taken as

$$2i\frac{\partial\Psi}{\partial t}+\frac{\partial^2\Psi}{\partial x^2}+2\Psi|\Psi|^2+ip_0\tanh(L-|x|)\frac{\partial^2\Psi}{\partial x^2}+ig\Psi|\Psi|^2+\mu\Psi\frac{\partial\left(|\Psi|^2\right)}{\partial x}=0. \qquad (27)$$

Simulations of Eq. (27) were carried out by means of the standard Maple package using the finite-difference (mesh) method.

For $\sigma\equiv 2(g/p_0)-1=1$, i.e., $p_0=g$, which was identified above as the FP existence threshold, predicted by the analytical approximation, we simulated Eq. (27) with parameters

$$p_0=g=1/10,\ \mu=1/10,\ L=50. \qquad (28)$$

In this case, Eq. (22) predicts the with coordinates $A_*=5\sqrt{3}/6\approx 1.44$ and $k_*=5/6\approx 0.83$. Numerical simulations were performed with quasi-soliton initial conditions

$$\Psi(x,0)=A(0)\mathrm{sech}[A(0)x]\exp[i(5/6)x], \qquad (29)$$

cf. ansatz (15) adopted above for the analytical consideration. In input (29), the initial wavenumber $k(0)$ is taken equal to the FP's value, which is $k(0)=k_*\equiv 5/6$, for the parameters selected as per Eq. (28).

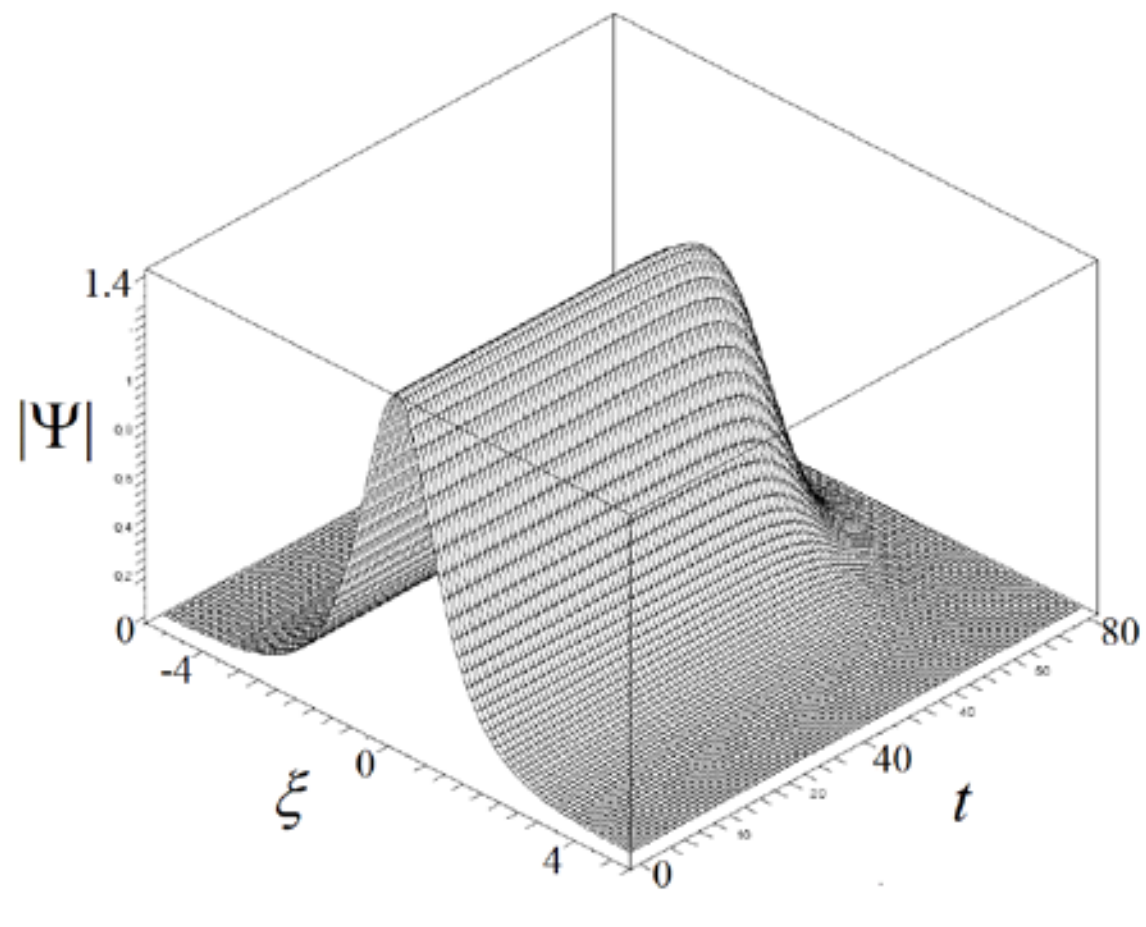

**Fig. 5.** Results of the simulations of Eq. (27) with parameters (28) and soliton-like input (29), with the initial amplitude $A(0)=A_*\equiv 5\sqrt{3}/6\approx 1.44$, $\xi=x-(5/6)t$. The quasi-soliton propagates stably and is then absorbed by the lossy boundary layer.

Simulations of Eq. (27) were performed for different initial amplitudes $A(0)$ in input (29) in the reference system moving with the equilibrium velocity $V=k_*\equiv 5/6$, this using the traveling coordinate $\xi = x - (5/6)t$. For the initial amplitude taken as the FP's value, $A(0) = A_* \equiv 5\sqrt{3}/6 \approx 1.44$, the direct simulations corroborate that the input pulse indeed propagates as a stable soliton, see Fig. 5. The soliton propagates keeping its profile and speed constant. Inputs which deviate from the equilibrium shape evolve toward the equilibrium (see below). Eventually, the wave packet, moving at the speed $V = k_* \equiv 5/6$ towards the boundary, hits the lossy region at time $t_* \approx L/V = 60$, and is absorbed there.

Further, Fig. 6 displays the results produced by the simulations for the input (29) with the initial amplitude different from the FP's value, *viz*., $A(0) = 1.2A_* \approx 1.73$ in (a), and $A(0) = 1.5A_* \approx 2.16$. In agreement with the prediction of the analytical approximation, in the former case the input propagates as a stably oscillating breather, which corresponds to the fact that the analytical approximation predicts that the FP is the stationary solution of the center type, and solution around it perform persistent oscillations, provided that the initial conditions belong to interval (25). Indeed, for the present values of the parameters, that interval amounts to

$$0.65A_* \approx 0.94 < A(0) < 1.26A_* \approx 1.82, \tag{30}$$

and the value $A(0) \approx 1.73$ corresponding to Fig. 6(a) belongs to the region (30). Eventually, the moving breather is absorbed by the lossy boundary layer.

On the other hand, the initial value $A(0) \approx 2.16$, which corresponds to Fig. 6(b), does not belong to the region (30), which explains the inference that the input decays to zero in this case, without hitting the absorptive boundary layer.

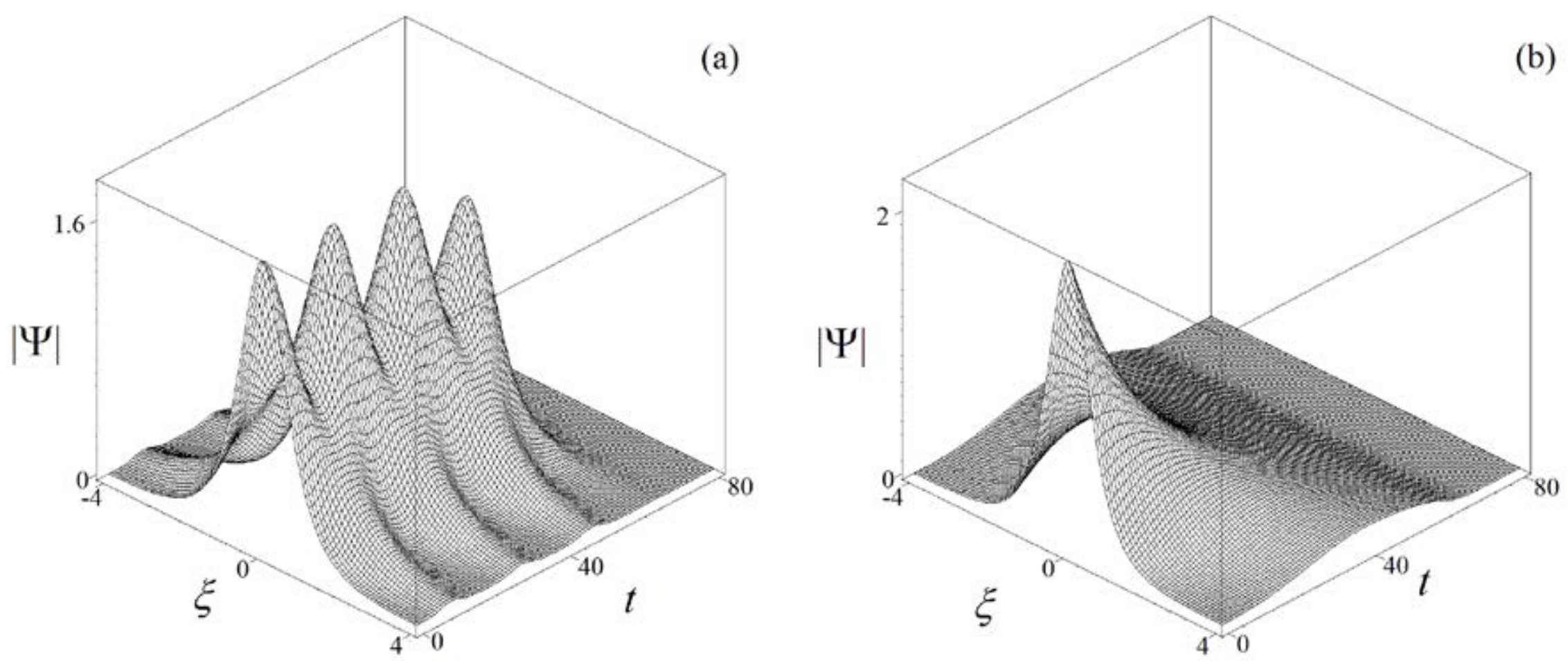

**Fig. 6.** Results of the simulations of Eq. (27) in the reference system moving with velocity $V = k_* \equiv 5/6$ $(\xi = x - (5/6)t)$ with parameters (28) and input (29)

with initial amplitudes different from the FP's value $A_* \approx 1.44$, namely, $A(0) = 1.2A_* \approx 1.73$ in (a), and $A(0) = 1.5A_* \approx 2.16$ in (b).

Next, we consider the case of

$$p_0 = 1/10,\ g = 1/8,\ \mu = 1/10, \qquad (31)$$

for which $\sigma \equiv 2(g/p_0) - 1 \equiv 3/2$. In this case, Eq. (22) predicts the FP's amplitude $A_* \equiv 5\sqrt{2}/4 \approx 1.78$. As shown above, the analytical approximation predicts the FP in the form of the stable spiral for $\sigma > 1$. The respective FP's coordinates, as given by Eq. (22), are $A_* = 5\sqrt{2}/4 \approx 1.78$ и $k_* = 5/4 \approx 1.25$, while the respective input is again selected as per Eq. (29).

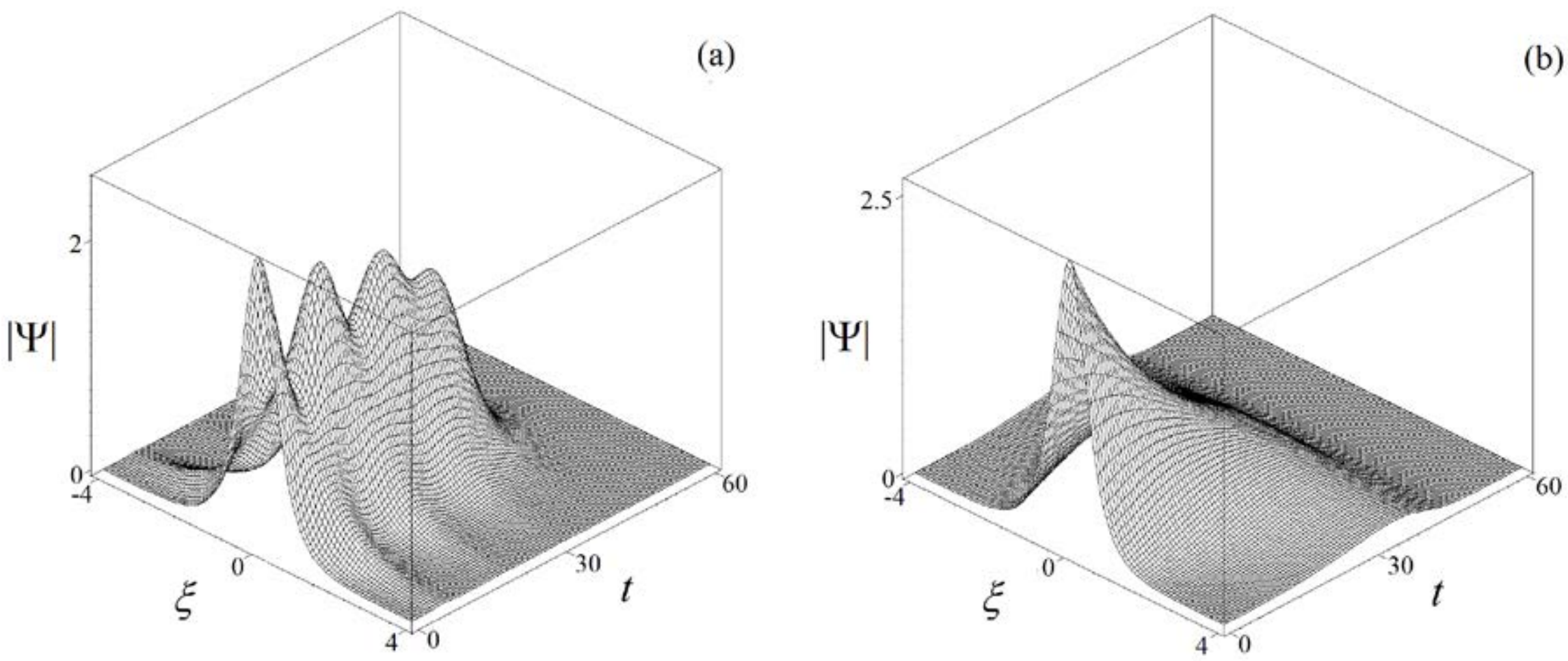

**Fig. 7.** Results of the simulations of Eq. (27) in the reference system moving with velocity $V = k_* \equiv 5/4$ $(\xi = x - (5/4)t)$, with parameter values (31) and input (29) with initial amplitudes $A(0) = 1.4A_* \approx 2.49$ in (a), and $A(0) = 1.5A_* \approx 2.65$ in (b).

In the present case, results of the simulations of Eq. (27) are reported in Fig. 7, for initial amplitudes $A(0) = 1.4A_* \approx 2.49$ in (a), and $A(0) = 1.5A_* \approx 2.65$ in (b). In agreement with the prediction of the analytical approximation, Fig. 7(a) demonstrates attraction of the non-equilibrium input to the FP in interval (26) of the initial values of the amplitude. In terms of the current notation, this interval is

$$0 < A(0) \le 1.42A_* \approx 2.53. \qquad (32)$$

Similar to the cases shown in Figs. 5(a) and 6(a), the propagating wave packet, moving at the speed $V = k_* \equiv 5/4$, eventually hits the lossy boundary layer at time $t_* \approx L/V = 40$, and is then quickly absorbed. On the other hand, the initial amplitude $A(0) \approx 2.65$ used in Fig. 7(b) does not belong to interval (32), therefore the latter figure demonstrates decay of the input.

It is relevant to mention that the analytical approximation, developed above in the form of Eq. (22), produces an indefinite result, in the form of 0/0, in the special case of $\mu = 0$ and $p_0 = 2g$ (i.e., $\sigma = 0$). In this case, Eq. (27) takes the form of

$$2i\frac{\partial\Psi}{\partial t} + \frac{\partial^2\Psi}{\partial x^2} + 2\Psi|\Psi|^2 + ip_0 \tanh(L - |x|)\frac{\partial^2\Psi}{\partial x^2} + i\frac{p_0}{2}\Psi|\Psi|^2 = 0. \quad (33)$$

Direct simulations of Eq. (33) demonstrate that, unlike the full equation (27), it does not maintain steady propagation of quasi-solitons. For example, with parameters $p_0 = 1/10$, $L = 50$ and input

$$\Psi(x,0) = (1/2)\operatorname{sech}(x)\exp(ikx), \quad (34)$$

with different values of $k$, the simulations produce the results plotted in Fig. 8, with panels (a) and (b) corresponding to $k = 3$ and $k = 5$, respectively. It is seen that the wave pulse grows under the action of the unbalanced gain and moves with acceleration. Eventually, it quickly hits the lossy boundary layer at time $t_* \approx 8$ for $k = 3$, and $t_* \approx 4$ for $k = 5$, and is then quickly absorbed.

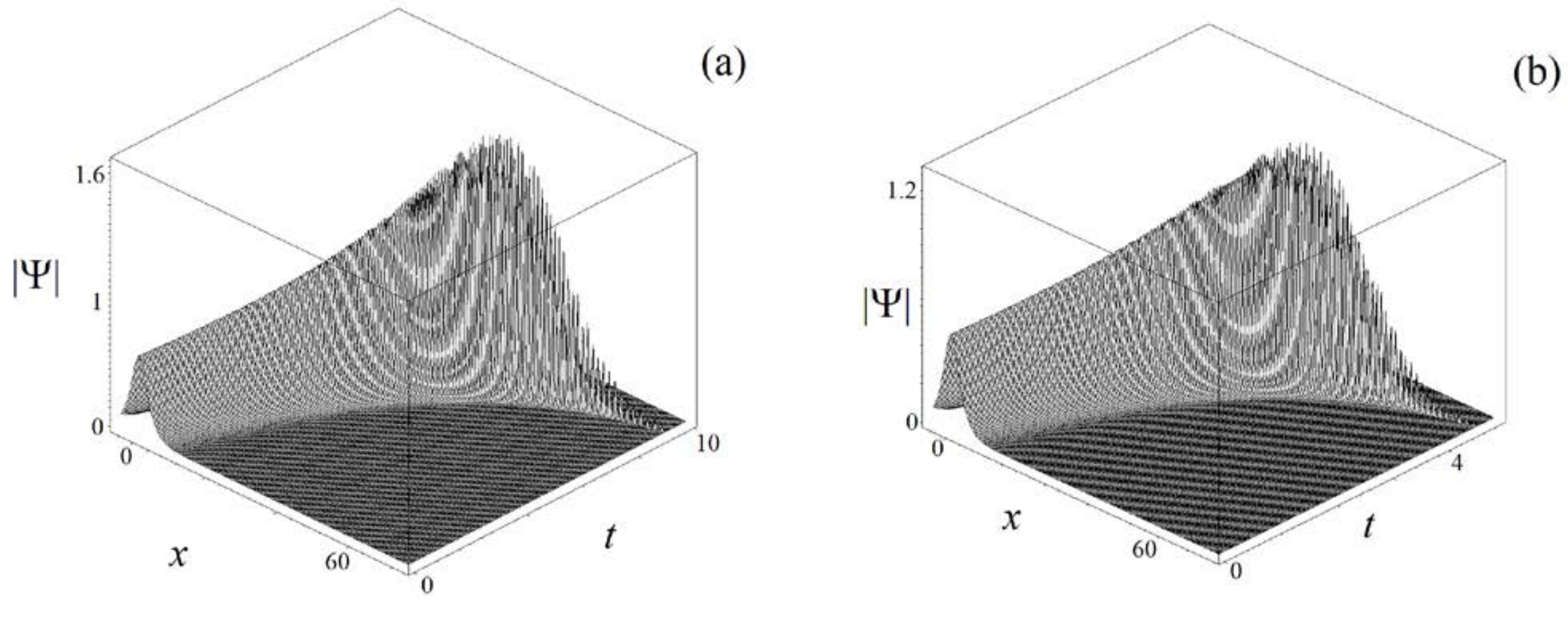

**Fig. 8.** The result of the simulations of Eq. (33) with parameters $p_0 = 1/10$, $L = 50$ and input (34) with $k = 3$ (a) and $k = 5$ (b).

## 5. Conclusion

We have considered the dynamics of the model which is based on the nonconservative NSLE. It includes the cubic pseudo-SRS (simulated Raman scattering) term, cubic nonlinear loss, and the anti-diffusion gain. The nonconservative NSLE can be derived from the ZS (Zakharov's system) in which the anti-diffusion gain in the HF (high-frequency) component is produced by the expansion with respect to deviation of the wavenumber around the point of the maximum loss (on the contrary to the usual models, in which the expansion is performed around the minimum-loss point). The LF (low-frequency) components of the ZS field includes the usual viscosity. The convective instability, naturally produced by the anti-diffusion gain, is suppressed by absorptive boundaries in the finite-length setting. Steadily moving soliton states of the model are predicted, as FPs (fixed points), by means of the analytical approximation, based in the balance equations for the system's norm and momentum. An analytical solution is also obtained for nonstationary solutions of the corresponding approximate system, and the FP's attraction basin is identified too. The analytical prediction is corroborated by direct simulations of the underlying nonconservative NSLE with the absorptive boundary conditions. As an extension of the analysis, it may be interesting to consider collisions between the solitons moving with different velocities.

## Statements and Declarations

**Funding:** The authors did not receive support from any organization for the submitted work.

**Availability of data and material:** No datasets were generated or analysed during the current study

**Competing Interests:** The work was carried out entirely on the initiative of the authors, as a natural continuation of joint scientific research on the topic of nonlinear waves, solely driven by scientific interests, without any financial support from any party or external source.

**Author contributions:** All authors contributed to the study conception and design. Material preparation, data collection and analysis were performed by B.A. Malomed and E.M. Gromov. The first draft of the manuscript was written by E.M. Gromov and all authors commented on previous versions of the manuscript. All authors read and approved the final manuscript.